\documentclass{PoS}
\usepackage{dsfont}

\title{The critical endpoint in the 2d U(1) gauge-Higgs model at topological angle $\theta=\pi$}

\ShortTitle{The critical endpoint in the 2d U(1) gauge-Higgs model at topological angle $\theta=\pi$}

\author{\speaker{Daniel G\"oschl} \thanks{ This work is supported by the 
Austrian Science Fund FWF, grant I 2886-N27 and the FWF
DK ''Hadrons in Vacuum, Nuclei and Stars'', grant W-1203. 
Parts of the computational results presented have been achieved using the Vienna Scientific Cluster (VSC).}\\
        Institute of Physics, University of Graz, Austria\\
        E-mail: \email{daniel.goeschl@uni-graz.at}}

\author{Christof Gattringer\\
        Institute of Physics, University of Graz, Austria\\
        E-mail: \email{christof.gattringer@uni-graz.at}}

\author{Tin Sulejmanpasic\\
        Philippe Meyer Institute, Physics Department, \'Ecole Normale Sup\'erieure,\newline 
        PSL Research University, 24 rue Lhomond, F-75231 Paris Cedex 05, France\\
        E-mail: \email{tin2019@gmail.com}}

\abstract{We study 2d U(1) gauge Higgs systems with a $\theta$-term. For properly discretizing the topological 
charge as an integer we introduce a mixed group- and algebra-valued discretization (MGA scheme) for the gauge 
fields, such that the charge conjugation symmetry  at $\theta = \pi$ is implemented exactly. The complex action 
problem from the $\theta$-term is overcome by exactly mapping the partition sum to a worldline/worldsheet 
representation. Using Monte Carlo simulation of the worldline/worldsheet representation we study the system at 
$\theta = \pi$ and show that as a function of the mass parameter the system undergoes a phase transition. 
Determining the critical exponents from a finite size scaling analysis we show that the transition is in the 2d Ising 
universality class. We furthermore study the U(1) gauge Higgs systems at $\theta = \pi$ also with charge 2 matter 
fields, where an additional $\mathds{Z}_2$ symmetry is expected to alter the phase structure. Our results indicate that 
for charge 2 a true phase transition is absent and only a rapid crossover separates the large and small mass regions.}

\FullConference{The 36th Annual International Symposium on Lattice Field Theory - LATTICE2018\\
		22-28 July, 2018\\
		Michigan State University, East Lansing, Michigan, USA.}

\begin{document}

\section{Introduction}

One of the many exciting features of quantum field theories is that they allow for the introduction of topological
terms. Such terms may alter the symmetry structure of the system and consequently give 
rise to new physics. Since topological terms capture global properties of a system, non-perturbative techniques 
are needed to study the corresponding physics.

In principle lattice field theory is a suitable non-perturbative approach, as long as two key challenges can be 
overcome: The lattice discretization of topological terms is not straightforward due to the absence of the notion 
of smoothness of the fields. The second challenge is the fact that topological terms typically give rise to a
complex action, such that the Boltzmann factor cannot be used as a probability weight in a Monte Carlo simulation
(''complex action problem'').

Here we sketch an approach that solves the two challenges for 2d U(1) gauge Higgs models, 
which constitute a class of systems interesting as toy models for high energy physics, as well as in condensed matter
theory. The approach is based on a mixed group- and algebra-valued lattice discretization of the gauge fields 
(MGA discretization) that correctly implements
the symmetries related to the U(1) $\theta$-term and combine this with an exact mapping to a worldline/worldsheet
representation that solves the  complex action problem.

To be more specific, the simplest model we study with the new discretization approach is the 2d U(1) gauge 
Higgs model which in the continuum is described by the action
\begin{equation}
S \; = \; \int_{\mathds T^2} d^2x \left(|D_\mu\phi|^2 \, + \, m^2 |\phi|^2 \, + \, \lambda |\phi|^4 \,+ \, 
\frac{1}{2\,e^2}F_{12}^{\,\;2} \, + \, 
i \, \theta \, \frac{1}{2\pi} F_{12}\right) \; .
\label{S_cont}
\end{equation}
$\phi(x) \in \mathds{C}$ denotes the charged scalar field and $A_\mu(x) \in \mathds{R}$ the U(1) gauge field.
$D_\mu=\partial_\mu+i\,A_\mu$ is the U(1) covariant derivative and $F_{12}=\partial_1A_2-\partial_2A_1$ denotes 
the field strength tensor. $m$ is the bare mass, $\lambda$ the coupling for the quartic self-interaction, $e$ the electric 
charge and $\theta$ the topological angle. We study the theory on a 2-torus $\mathds T^2$, where 
(for sufficiently smooth fields) the flux of $F_{12}$ is quantized in integer units of $2\pi$, such that the topological 
charge $Q_{top} = \frac{1}{2\pi} \int d^2x F_{12}$ is an integer.
The partition function of the system is given by 
$Z  = \int \! D[\phi] \int \! D[A] \; e^{-S[\phi, \, A,\theta]}$.

Charge conjugation transforms the fields as $\phi(x) \rightarrow \phi(x)^*$, $A_\mu(x) \rightarrow - A_\mu(x)$, and
the field strength changes its sign $F_{12} \rightarrow -F_{12}$ under this transformation.  The gauge field and the 
matter field parts of the action remain invariant, while the topological charge $Q_{top}$ changes its sign. 
Since $Q_{top}$ is an integer, charge conjugation is a symmetry not only at the trivial value $\theta = 0$, but also at 
$\theta = \pi$.  This $\mathds{Z}_2$ symmetry is expected to be intact for small mass $m$, but broken at large $m$. 
It has been conjectured \cite{Komargodski:2017dmc}
that at intermediate mass values there is a second order phase transition in the 
universality class of the 2d Ising model. Establishing this conjecture from an ab-initio lattice calculation was one of the 
goals of this project, where we use the new MGA scheme to discretize the 
topological charge as an integer (for more details and motivation see \cite{Gattringer:2018dlw}). 
Thus charge conjugation symmetry at $\theta = \pi$ is implemented exactly and a mapping to a 
worldline/worldsheet representation with only real and positive weights solves the complex action problem. 

Also more general systems of the type (\ref{S_cont}) are of interest and need to be studied in 
a suitable lattice formulation -- in particular matter fields with a higher charge and also generalizations
to more than two flavors. New symmetries appear and the phase structure changes. As a preview to future work
for this type of systems we present first results for the U(1) gauge Higgs model at $\theta = \pi$ but with scalar fields
of charge 2.

\section{Mixed group- and algebra-valued lattice discretization (MGA discretization)}

We begin the discussion of the MGA discretization with the lattice action $S_H[\phi,U]$ for the 
matter field (the lattice spacing $a$ is set to $a = 1$), 
\begin{equation}
S_M[\phi,U] \; = \; \sum_{x\in\Lambda}\left[ M |\phi_x|^2 + \lambda |\phi_x|^4 - \sum_{\mu=1}^{2} 
\Big(\phi_{x}^{\ast} U_{x,\mu} \phi_{x+\hat{\mu}}+c.c.\Big) \right] \; ,
\label{S_Higgs}
\end{equation}
where the mass parameter $M$ is related to the bare mass $m$ via $M = 4 + m^2$. The gauge fields
couple via the group-valued link variables $U_{x,\mu} \in$ U(1). We parameterize the link variables
in the form $U_{x,\mu} = e^{\, i \, A_{x,\mu}}$, with the algebra-valued lattice gauge fields 
$A_{x,\mu} \in \mathds{R}$. 

The gauge field action and the $\theta$-term will now be discretized with the algebra-valued fields $A_{x,\mu}$. 
For this step we note that due to the use of the group-valued link variables $U_{x,\mu}$, the matter field action 
(\ref{S_Higgs}) is invariant under the shifts
\begin{equation}
A_{x,\mu} \; \rightarrow \; A_{x,\mu} \; + \; 2 \pi \, k_{x,\mu} \quad , \qquad k_{x,\mu} \; \in \; \mathds{Z} \; ,
\label{shifts}
\end{equation}
and the discretization of the gauge field action and topological term will have to take into account this 
invariance. Using the group-valued fields $A_{x,\mu}$, a natural definition of the field strength is
$F_{x,12} = A_{x+\hat{1},2} - A_{x,2} -A_{x+\hat{2},1}  + A_{x,1} \equiv F_x$, that is assigned
to the plaquettes of the lattice, which in 2d can be labeled by the coordinate $x$ of the lower left corner
of the plaquette. However, this definition of $F_x$ is not invariant under the shifts (\ref{shifts}), where it transforms as
$F_x \rightarrow F_x + 2 \pi ( k_{x+\hat{1},2} - k_{x,2} -k_{x+\hat{2},1}  + k_{x,1} )$, i.e., 
$F_x$ is shifted by multiples of $2 \pi$. In order to recover invariance we implement the following discretization
strategy: The continuum field strength $F_{12}(x)$ is replaced by 
$A_{x+\hat{1},2}  -  A_{x,2}  - A_{x+\hat{2},1} +  A_{x,1} +  2 \pi \, n_x$,
and the plaquette-based auxiliary variables $n_x$ are summed over all integers. Obviously this construction 
recovers the invariance under the shifts (\ref{shifts}). 

Using this prescription we can write the Boltzmann factor $B_G[A]$ for the algebra-valued gauge field, which 
takes into account the gauge field action and the $\theta$-term as ($\beta \equiv 1/e^2$)
\begin{equation}
B_G [A]  \; = \;  \prod_{x \in \Lambda} \, \sum_{n_x \in \mathds{Z}} 
\, e^{ \, - \frac{\beta}{2} ( F_{x}  +  2 \pi \, n_x)^2 \, - \, i \, \frac{\theta}{2\pi} \, (F_{x}  +  2 \pi \, n_x)} \; = \;  
\sum_{\{ n \}} e^{ \, -  \frac{\beta}{2} \sum_x ( F_{x}  +  2 \pi \, n_x)^2 \, - \, i \, \theta \sum_x n_x} \;  . 
\label{BG}
\end{equation} 
In the second step we have introduced $\sum_{\{n\}} \equiv \prod_x \sum_{n_x \in \mathds{Z}}$ for the
sum over all configurations of the auxiliary variables $n_x$. Furthermore, in that step we have written the product
$\prod_x$ as a sum over $x$ in the exponent and used the fact that $\sum_x F_x = 0$ for a lattice with periodic 
boundary conditions. Consequently this step identifies the topological charge as $Q_{top} = \sum_x n_x$, which
obviously is an integer, such that the first criterion for our lattice discretization is obeyed. We remark that at 
$\theta = 0$ our Boltzmann factor (\ref{BG}) reduces to the well known Villain form \cite{Villain:1974ir}.

Our MGA lattice discretization is completed by combining the building blocks into 
the lattice path integral for the partition sum, 
\begin{equation}
Z \; = \; \int \! D[A] \; B_G[A] \; Z_M[U]  \quad , \qquad Z_M[U] \; = \; \int \,D[\phi] \ e^{-S_M[\phi,U]} \; ,
\label{Z_lat}
\end{equation}
with the corresponding measures defined as $\int \! D[\phi] = \prod_x \int_{\mathds{C}}  \frac{d\phi_x}{2\pi}$ and 
$\int \! D[A] = \prod_{x,\mu} \int_{-\pi}^{\pi} \frac{dA_{x,\mu}}{2\pi}$.

We conclude our discussion by showing that at $\theta = \pi$ 
the partition sum (\ref{Z_lat}) with matter field action (\ref{S_Higgs}) and Boltzmann factor (\ref{BG}) is indeed
invariant under charge conjugation. Charge conjugation is implemented as in the continuum via $\phi_x \rightarrow 
\phi_x^*$ and $A_{x,\mu} \rightarrow - A_{x,\mu}$. The latter implies $U_{x,\mu} \rightarrow U_{x,\mu}^*$, which 
together with $\phi_x \rightarrow \phi_x^*$ ensures the invariance of $Z_M[U]$. For the field strength 
we find again $F_x \rightarrow - F_x$, which in the quadratic term $(F_x + 2\pi n_x)^2$ of the gauge field
Boltzmann factor (\ref{BG}) can be compensated by transforming also the auxiliary variables via 
$n_x \rightarrow - n_x$. This implies for the transformation of the Boltzmann factor with the topological charge
$e^{- i \theta \sum_x n_x} \rightarrow e^{+ i \theta \sum_x n_x}$ which is invariant for $\theta = \pi$ (and $\theta = 0$).

\section{Representation with worldlines and worldsheets}

Having found an approach that discretizes the topological charge as an integer and thus exactly implements the 
charge conjugation symmetry at $\theta = \pi$, we now come to solving the complex action problem by transforming 
the partition sum to a worldline/worldsheet representation. This transformation has been discussed for U(1) gauge 
Higgs systems in, e.g., \cite{Mercado:2013yta,Mercado:2013ola} and for the MGA discretization is derived in detail 
in \cite{Gattringer:2018dlw}. Thus we here only provide a short sketch of the derivation and mainly discuss the final 
form we use for the numerical simulation.

The partition function $Z_M [U]$ for the matter field in a background $U$ of the compact link variables is clearly 
a gauge invariant functional. The only gauge invariant quantities one can form with the link variables
correspond to products of link variables $U_{x,\mu}$ placed along closed loops. Such closed loops can be
described by integer valued flux variables $j_{x,\mu} \in \mathds{Z}$ assigned to the links of the lattice. The value
$j_{x,\mu}$ for the flux indicates how often a link is run through by loops, where negative values correspond to 
fluxes in negative direction. The requirement that the loops are closed is implemented by enforcing zero divergence
$\vec{\nabla}\cdot \vec{j}_x \; = \; \sum_\mu [j_{x,\mu} - j_{x-\hat{\mu},\mu}] \; = \; 0$ at every site of the lattice. 
The contribution of a link variable $U_{x,\mu}$ is then simply given by $( U_{x,\mu}) ^{\, j_{x,\mu}}$. Thus
the matter field partition sum can be written as
\begin{equation}
Z_M [U] \; = \; \sum_{\{j\}} W_M[j] \, \prod_x \delta\left(\vec{\nabla}\cdot \vec{j}_x\right) 
\, \prod_{x,\mu} \, ( U_{x,\mu}) ^{\, j_{x,\mu}} 
\; = \; \sum_{\{j\}} W_H[j] \, \prod_x \delta\left(\vec{\nabla}\cdot \vec{j}_x\right) 
\, \prod_{x,\mu} \, e^{\, i \, A_{x,\mu} \; j_{x,\mu}} \; ,
\label{Z_H_dual}
\end{equation}
where we have defined $\sum_{\{j\}} = \prod_{x,\mu} \sum_{j_{x,\mu} \in \mathds{Z}}$ to denote the sum over all 
configurations of the flux variables $j_{x,\mu}$. The zero divergence condition is implemented by a product 
of Kronecker deltas (here denoted by $\delta(n) \equiv \delta_{n,0}$) at all sites. The gauge field dependence is 
the product of $( U_{x,\mu}) ^{\, j_{x,\mu}}$ over all links, where in the second step in (\ref{Z_H_dual}) we have 
already inserted $U_{x,\mu} = e^{ \, i A_{x,\mu}}$.

The configurations of the flux variables $j_{x,\mu}$ come with real and positive weight factors $W_H[j]$ that can be
determined by an expansion of the nearest neighbor Boltzmann factors of the matter fields and a subsequent 
integration over the matter fields $\phi_x$ (see \cite{Gattringer:2018dlw,Mercado:2013yta,Mercado:2013ola} 
for their derivation).  

The next step for finding the worldline/worldsheet representation is to represent the gauge field Boltzmann factor
$B_G[A]$ by its Fourier transform. Since the Boltzmann factor $B_G[A]$ is $2\pi$-periodic in the $F_x$, the 
Fourier representation will depend on the gauge fields in the form $\prod_x e^{\,  i \, F_x \, p_x}$ where the 
Fourier modes $p_x \in \mathds{Z}$ are assigned to the plaquettes and referred to as ''plaquette occupation
numbers''. Since the Boltzmann factor (\ref{BG}) is Gaussian, the Fourier transforms can be computed 
in closed form, such that one finds (see, e.g., \cite{Gattringer:2018dlw} for details) 
\begin{equation}
B_G[A] \; = \, \sum_{\{ p \}} e^{ -\frac{1}{2 \beta} \sum_x \big( p_x + \frac{\theta}{2\pi} \big)^2} \!\! \prod_x 
e^{\,  i \, F_x \, p_x} \; = \,  \sum_{\{ p \}} e^{ -\frac{1}{2 \beta} \sum_x \big( p_x + \frac{\theta}{2\pi} \big)^2} \!\!
\prod_x \, e^{\,  i A_{x,1} [ p_x - p_{x-\hat{2}}] } \, e^{\,  - i A_{x,2} [p_x - p_{x-\hat{1}}]}.
\label{BG_dual}
\end{equation}
Here $\sum_{\{ p \}} =  \prod_{x} \sum_{p_{x} \in \mathds{Z}}$ denotes the sum over all configurations of the 
plaquette occupation numbers. In the second step we have reorganized the gauge field
dependence already in terms of the non-compact gauge fields $A_{x,\mu}$. The configurations of the
plaquette occupation numbers come with a Gaussian weight, and the topological angle 
determines the position of the center of the Gaussian. 

In a final step we integrate over the gauge fields $A_{x,\mu}$ at all links. The factors $e^{ \, i A_{x,\mu} \, j_{x,\mu}}$
from the matter field partition sum $Z_M[U]$ and the factors 
$e^{\,  i A_{x,1} [ p_x - p_{x-\hat{2}}] } \, e^{\,  - i A_{x,2} [p_x - p_{x-\hat{1}}]}$ from the Boltzmann factor $B_G[A]$
are linked together by this integration to a new set of constraints that now live on the links of the lattice. We thus
obtain the final form of the worldline/worldsheet representation,
\begin{equation}
Z \; =  \; \sum_{\{p,j\}} e^{\, - \frac{1}{2 \beta} \sum_{x} \, \big( p_{x} \, + \, 
\frac{\theta}{2\pi} \big)^2 } \; W_H[j] \; \; 
\prod_x \delta \big( \vec{\nabla} \cdot \vec{j}_x \big) \; 
 \delta  \left(  j_{x,1}  +  p_{x} - p_{x-\widehat{2}} \right) \;  
 \delta  \left(  j_{x,2}  -   p_{x} + p_{x-\widehat{1}}  \right) .
\label{Z_dual}
\end{equation}
The partition function is a sum over all configurations of the plaquette occupation numbers $p_x$ and the flux 
variables $j_{x,\nu}$. They come with real and positive weight factors and two types of constraints: The zero 
divergence constraint that ensures conservation of the matter flux, as well as link-based constraints that enforce
the vanishing combined flux from the matter flux $j_{x,\mu}$ on a link and the neighboring plaquettes that contain 
the link. One finds that admissible configurations are closed loops of matter flux which are filled with occupied 
plaquettes, i.e., plaquettes with $p_x \neq 0$ form patches (2d surfaces) that are bounded by matter flux. 
Since all weights in (\ref{Z_dual}) are real and positive, numerical simulations can be done in terms of the 
flux variables $j_{x,\mu}$ and the plaquette occupation numbers $p_x$ and the complex action problem is solved. 
Suitable update schemes that properly take into account the constraints are discussed in \cite{Mercado:2013yta}. 

The observables we consider here are the expectation value of the topological charge density $q = Q_{top}/V$, where 
$V$ is the volume of the lattice and the corresponding susceptibility $\chi_t$. They correspond to the first and second 
derivative of $\ln Z$ with respect to the topological angle $\theta$. We can evaluate these derivatives using the dual 
form (\ref{Z_dual}) of the partition sum and obtain the two observables in terms of the first and second moments of 
the plaquette occupation numbers. For details of this derivation and the technical aspects of the numerical 
simulation we refer to \cite{Gattringer:2018dlw}. For a lattice study of the same system based on the Wilson
discretization see \cite{Gattringer:2015baa}.

\section{Numerical results}

\begin{figure}[t!]
	\centering
	\includegraphics[width=126mm,clip]{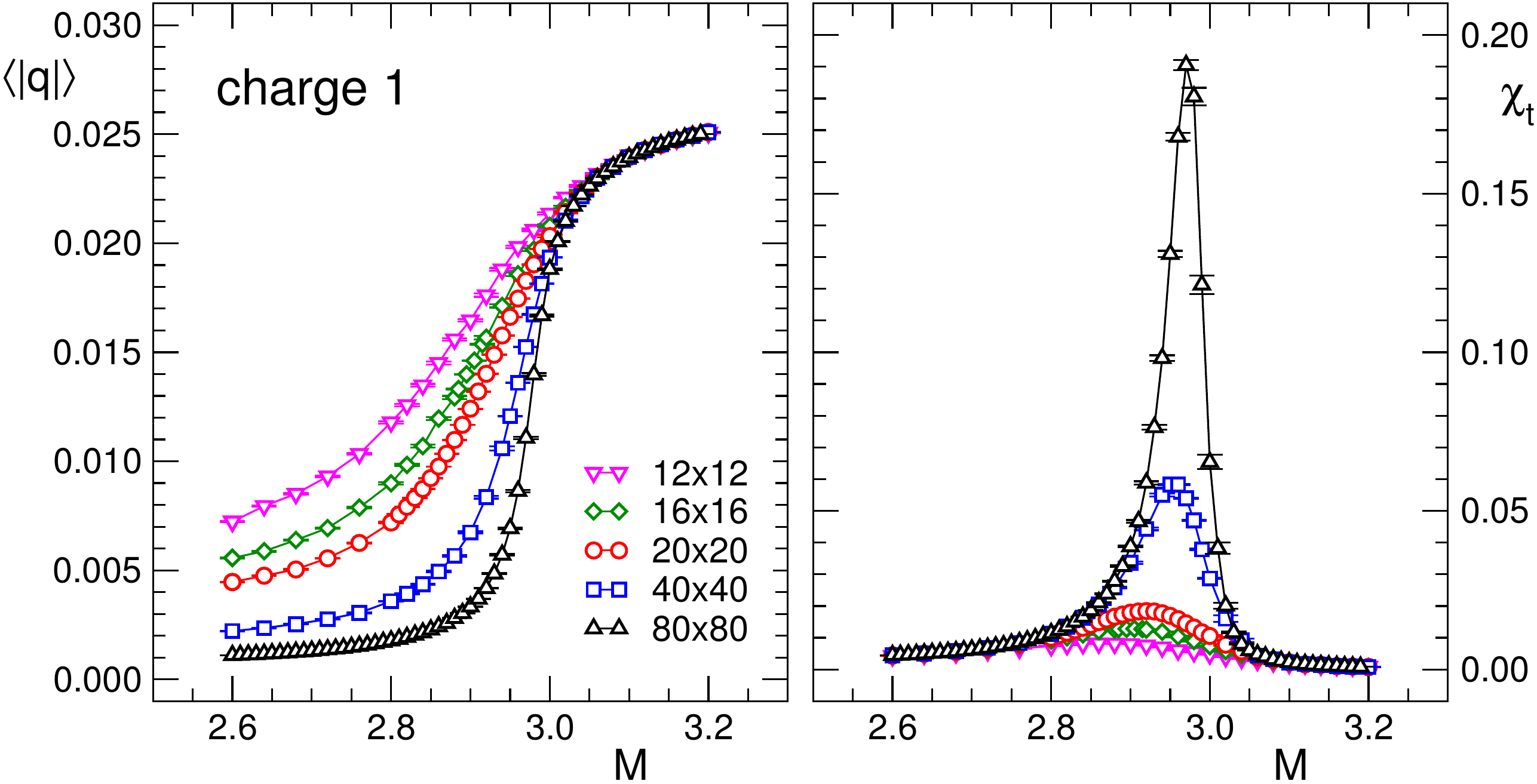}
		\caption{The topological charge density $\langle | q | \rangle$ and the susceptibility  $\chi_t$
		at $\theta = \pi$ for different volumes. 
		We show the charge 1 results for $\beta = 3.0$, $\lambda = 0.5$ 
		and plot the observables as a function of $M = 4 + m^2$.}
	\label{q_chi_vs_M_1}
\end{figure}

As we have outlined in the introduction, we aim at studying the system at $\theta = \pi$ as a function of the 
mass parameter. At small $M = 4 + m^2$ one expects that charge conjugation symmetry is intact, while
at large $M$ it should be broken \cite{Komargodski:2017dmc}. For some critical value $M_c$ a second order phase 
transition in the  2d Ising universality class is expected. A suitable infinite volume order parameter is the expectation 
value $\langle q \rangle$ of the topological charge ($q$ is odd under charge conjugation), which on a finite lattice
needs to be replaced by $\langle | q | \rangle$. 

In the left hand side plot of Fig.~\ref{q_chi_vs_M_1} we show $\langle | q | \rangle$ at $\theta = \pi$ 
as a function of $M$ and indeed we observe the transition from a symmetric phase with $\langle | q | \rangle = 0$ 
at small $M$ into a broken phase with $\langle | q | \rangle \neq 0$ at sufficiently large $M$. The corresponding 
susceptibility in the rhs.\ plot of Fig.~\ref{q_chi_vs_M_1} shows maxima that grow with the volume, which is an 
indication of a phase transition. 

\begin{figure}[b!]
	\centering
	\hspace*{2mm}
	\includegraphics[width=125mm,clip]{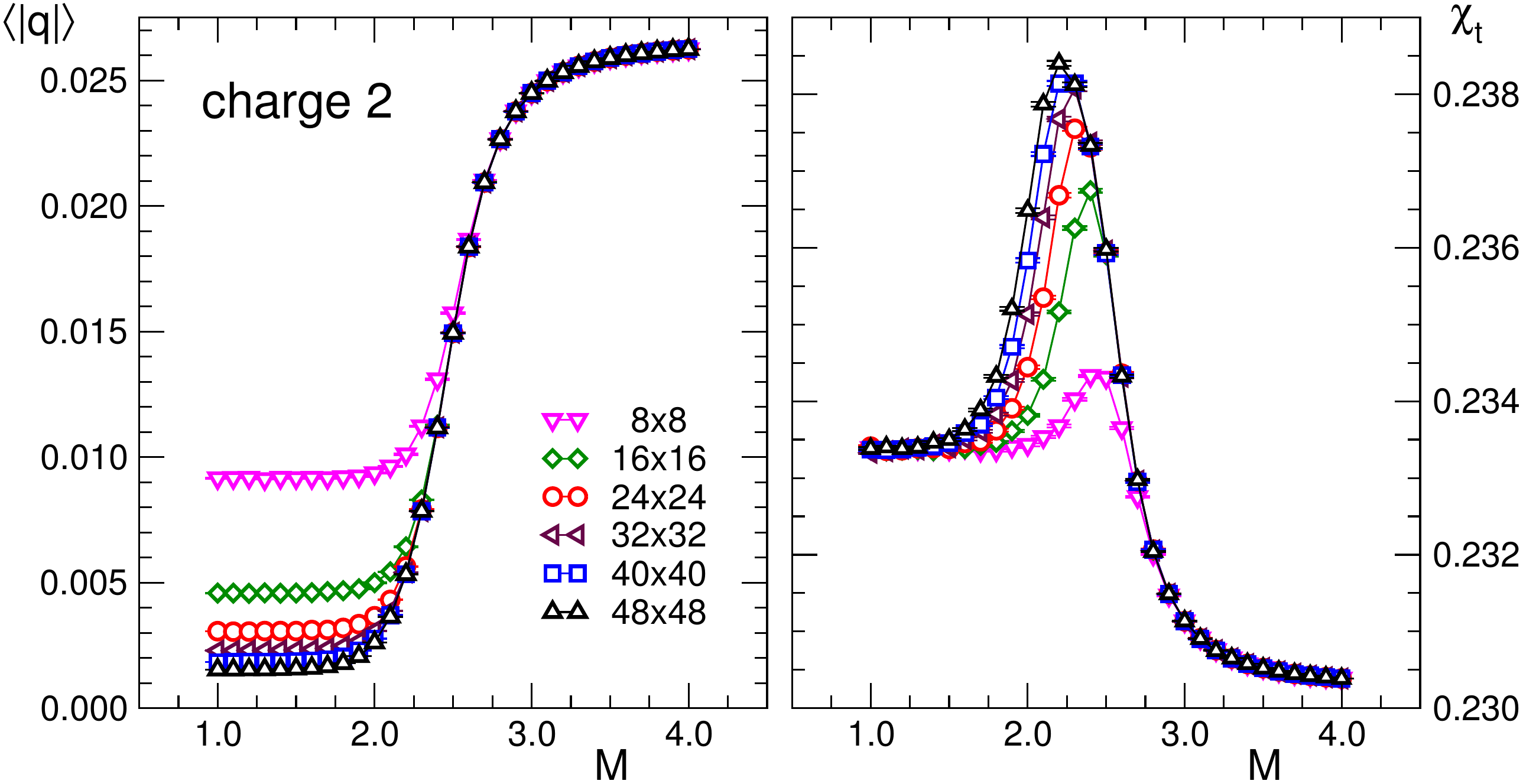}
	\caption{The topological charge density $\langle | q | \rangle$ and the susceptibility  $\chi_t$
		at $\theta = \pi$ for different volumes. 
		We show the charge 2 results for $\beta = 3.0$, $\lambda = 0.5$ 
		and plot the observables as a function of $M = 4 + m^2$.}
	\label{q_chi_vs_M_2}
\end{figure}

In \cite{Gattringer:2018dlw} we implemented a detailed finite volume scaling analysis in order to determine the 
critical exponents for the transition. The results are $\nu = 1.003(11)$, $\beta = 0.126(7)$ and
$\gamma = 1.73(7)$, which are in good agreement with the 2d Ising exponents $\nu = 1$, $\beta = 0.125$ and
$\gamma = 1.75$. Thus by combining our MGA discretization with the worldline/worldsheet
representation we were able to establish the conjectured critical point in the 2d Ising universality class.

We already remarked that in the future we will consider generalizations of the simple U(1) gauge Higgs system 
with a topological term we have studied so far. Models with more than one flavor or different charges have 
different symmetries and may have altered anomaly matching conditions 
\cite{Sulejmanpasic:2018upi,Tanizaki:2018xto,Gaiotto:2017yup} such that the phase structure will be changed.
As a first step towards this direction we here show results for the U(1) gauge Higgs model with scalar 
fields of charge 2. The MGA discretization proceeds as for charge 1, but now 
in the action (\ref{S_Higgs}) for the matter fields the link variables 
$U_{x,\mu} = e^{ \, i \, A_{x,\mu}}$ are replaced by $(U_{x,\mu})^2 = e^{ \, i \, 2 \, A_{x,\mu}}$. In this case
one has an additional $\mathds{Z}_2$ symmetry under $U_{x,\mu} \rightarrow - U_{x,\mu}$, which in terms 
of the $A_{x,\mu}$ is given by $A_{x,\mu} \rightarrow A_{x,\mu} + \pi$. 

While it looks like the charge 2 model is a simple rescaling of the charge 1 theory, the two are in fact different. Namely 
the charge 2 model has a $\mathds{Z}_2$ center symmetry with an order parameter being the charge 1 Wilson loop. 
There is a mixed anomaly between the $\mathds{Z}_2$ center symmetry and the charge conjugation symmetry. This is 
evident from the fact that gauging the $\mathds{Z}_2$ center turns the theory into the usual charge 1 theory at 
topological angle $\theta=\pi/2$, which does not have charge conjugation symmetry 
(see \cite{Sulejmanpasic:2018upi,Tanizaki:2018xto,Gaiotto:2017yup} 
and references therein for related discussions).

As a consequence the anomaly between the $\mathds{Z}_2$ center symmetry and charge conjugation symmetry must 
be saturated by breaking one or the other. Since the $\mathds{Z}_2$ center symmetry (a 1-form symmetry) typically 
does not break in 1+1 dimensions due to instanton effects (in analogy to how instantons restore a discrete ordinary 
0-form symmetry  in quantum mechanics), we cannot have a phase without charge conjugation symmetry breaking, 
and there should be no phase transition. 

In Fig.~\ref{q_chi_vs_M_2} we show our observables $\langle | q | \rangle$ and $\chi_t$ as a function of $M$ 
for the charge 2 case at $\theta = \pi$. Again we observe transitory behavior at some critical value of $M$ where
the expectation value $\langle | q | \rangle$ rises quickly. However, the volume dependence is different and
the rhs.\ plot  clearly shows the absence of volume scaling for the peaks of the topological susceptibility. 
This indicates that no critical behavior emerges in the charge 2 model and the small and large mass regions are 
only separated by a crossover. This observation provides a first example
for how the changed symmetry content may alter the pattern of a transition originating from the presence of 
a topological term.

\bibliographystyle{utphys}
\bibliography{bibliography}

\end{document}